\documentclass[10pt,conference]{IEEEtran}
\IEEEoverridecommandlockouts
\usepackage{cite}
\usepackage{amsmath,amssymb,amsfonts}
\usepackage{algorithmic}
\usepackage{graphicx}
\usepackage{textcomp}
\usepackage{xcolor}

\usepackage{multirow}
\usepackage{bm}
\usepackage{enumitem}
\def\BibTeX{{\rm B\kern-.05em{\sc i\kern-.025em b}\kern-.08em
    T\kern-.1667em\lower.7ex\hbox{E}\kern-.125emX}}
\begin{document}

\title{Improving Code Search with Hard Negative Sampling Based on Fine-tuning}
\author{
\IEEEauthorblockN{Hande Dong$^{\ast}$}
\IEEEauthorblockA{
\textit{International Digital Economy Academy}\\
Shenzhen, China \\
donghd66@gmail.com}
\and
\IEEEauthorblockN{Jiayi Lin$^{\ast}$ \thanks{$^{\ast}$Co-first author}}
\IEEEauthorblockA{
\textit{International Digital Economy Academy}\\
Shenzhen, China \\
jiayilin1024@gmail.com}
\and
\IEEEauthorblockN{Yanlin Wang$^{\dag}$ \thanks{$^{\dag}$Corresponding author}}
\IEEEauthorblockA{
\textit{Sun Yat-sen University}\\
Guangzhou, China \\
wangylin36@mail.sysu.edu.cn}
\and
\IEEEauthorblockN{Yichong Leng}
\IEEEauthorblockA{
\textit{University of Science and Technology of China}\\
Hefei, China \\
lyc123go@mail.ustc.edu.cn}
\and
\IEEEauthorblockN{Jiawei Chen}
\IEEEauthorblockA{
\textit{Zhejiang University}\\
Hangzhou, China \\
sleepyhunt@zju.edu.cn}
\and
\IEEEauthorblockN{Yutao Xie}
\IEEEauthorblockA{
\textit{International Digital Economy Academy}\\
Shenzhen, China \\
yutaoxie@idea.edu.cn}
}

\maketitle

\begin{abstract}
    Pre-trained code models have emerged as the state-of-the-art paradigm for code search tasks. 
    The paradigm involves pre-training the model on search-irrelevant tasks such as masked language modeling, 
    followed by the fine-tuning stage, which focuses on the search-relevant task. 
    The typical fine-tuning method is to employ a dual-encoder architecture 
    to encode semantic embeddings of query and code separately, and then calculate their similarity based on the embeddings.
    
    However, the typical dual-encoder architecture falls short in modeling token-level interactions 
    between query and code, which limits the capabilities of model. 
    To address this limitation, 
    we introduce a cross-encoder architecture for code search that jointly encodes the concatenation of query and code. 
    We further introduce a Retriever-Ranker (RR) framework that cascades the dual-encoder and cross-encoder to promote the efficiency of evaluation and online serving. 
    Moreover, we present a ranking-based hard negative sampling (PS) method to improve the ability of cross-encoder to distinguish hard negative codes, 
    which further enhances the cascaded RR framework.
    Experiments on four datasets using three code models demonstrate the superiority of our proposed method.  
    We have made the code available at https://github.com/DongHande/R2PS. 
\end{abstract}


\section{Introduction}
Code search aims to retrieve relevant code snippets from large code repositories for a given natural language query~\cite{DBLP:conf/icse/GuZ018, DBLP:conf/sigsoft/CambroneroLKS019, DBLP:journals/tkdd/LingWWPMXLWJ21}, 
which can improve the productivity of programmers. 
The core of code search is to predict whether queries are relevant to codes. 
Traditionally, classical information retrieval (IR) methods such as term frequency-inverse document frequency (TF-IDF) have been widely adopted for code search~\cite{DBLP:conf/icse/DiamantopoulosK18}. 
Specialized rules are often designed to extract features of codes~\cite{DBLP:journals/pacmpl/LuanYBS019}, 
and term frequencies are exploited to identify relevant queries and codes~\cite{DBLP:conf/msr/SismanK13}. 
However, traditional IR methods have some inherent limitations; 
for instance, the designed rules are hard to cover complex and implicit features about the matching between queries and codes 
and the term mismatch issue can limit the ability to discover useful codes. 

Recently, deep learning techniques, particularly pre-trained models, have demonstrated remarkable success in the code search task, 
thanks to their ability to learn complex and implicit features~\cite{DBLP:conf/icse/GuZ018, DBLP:conf/iclr/GuoRLFT0ZDSFTDC21}. 
The models employed in code search tasks generally undergo two stages: pre-training and fine-tuning. 
Pre-training involves training the model with universal pretext tasks to enhance its code understanding ability. 
These tasks include masked language modeling~\cite{DBLP:conf/emnlp/FengGTDFGS0LJZ20}, 
identifier prediction~\cite{DBLP:conf/iclr/GuoRLFT0ZDSFTDC21}, 
contrastive learning~\cite{DBLP:conf/acl/GuoLDW0022}, 
editing~\cite{DBLP:conf/kbse/ZhangP0LG22}, and more. 
Fine-tuning involves training the model with the search task to improve its code search ability. 
Specifically, the search task aims to maximize the relevance score between relevant query and code 
while minimizing the relevance score between irrelevant query and code. 
While previous research for code search primarily focuses on the pre-training stage, 
i.e., proposing various training tasks to enhance the model's code understanding ability, 
there has been little focus on the fine-tuning stage. 
In this paper, we mainly focuses on improving the fine-tuning stage for code search. 

\begin{figure}[t]
    \centering
    \includegraphics[width=0.43\textwidth]{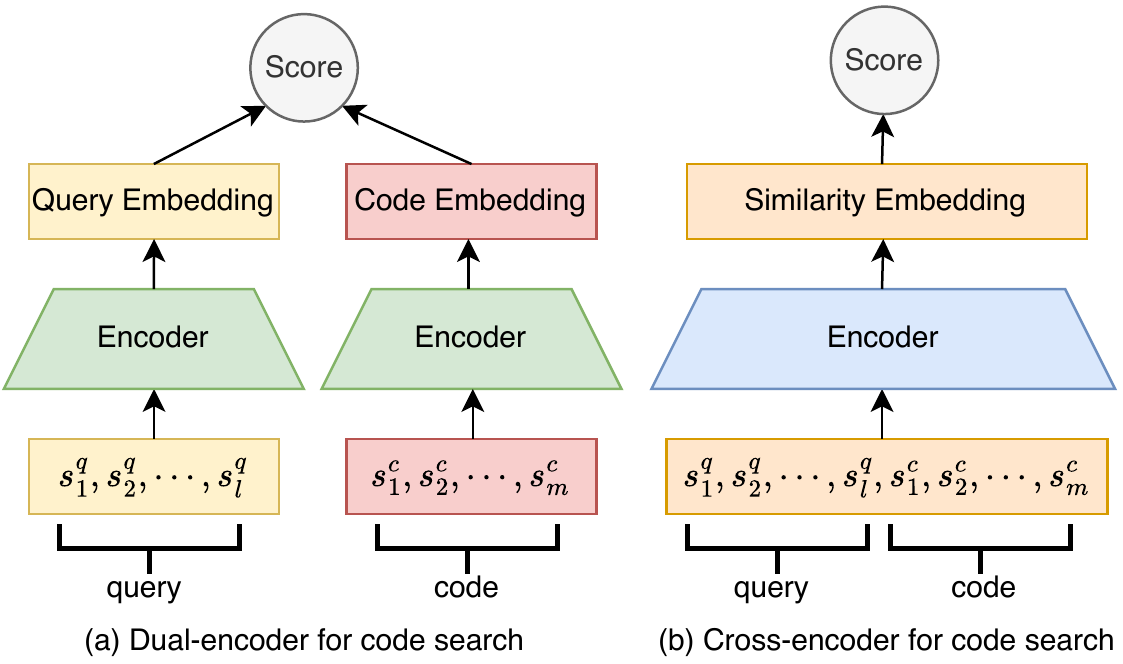}
    \caption{The dual-encoder architecture and the cross-encoder architecture for code search. 
    $s^q_1,s^q_2, \cdots,s^q_l$ denotes the token sequence of query $q$, and $s^c_1,s^c_2, \cdots,s^c_m$ denotes the token sequence of code $c$. 
    (a) In dual-encoder, we input the token sequence of the query and code separately; (b) In cross-encoder, we input the token sequence concatenation of the query and the code. }
    \label{intro_model}
\end{figure}

\IEEEpubidadjcol

During the fine-tune stage, models typically employ a dual-encoder architecture for code search, as shown in Figure~\ref{intro_model}(a). 
This architecture separately encodes the query and code to obtain their embeddings as semantic representations 
and then computes the relevance score between them based on these embeddings~\cite{DBLP:conf/acl/GuoLDW0022, DBLP:conf/kbse/ZhangP0LG22, DBLP:conf/icse/NiuL0GH022}. 
However, while the dual-encoder can fully model the token-level interactions 
within the query and code separately using self-attention mechanisms~\cite{DBLP:conf/cikm/GuoFAC16, DBLP:conf/sigir/XiongDCLP17}, 
it cannot model the token-level interactions cross query and code tokens. 
As a result, the dual-encoder's capability to learn fine-grained interactions between query and code tokens is limited.

In this paper, we propose Retriever and Ranker with Ranking-based Hard Negative Sampling method (R2PS) for code search, 
which is a patch designed to be used with various code pre-trained models. 
Compared to previous work that fine-tunes code pre-trained models with the dual-encoder architecture, R2PS has the following three advantages:

\paragraph{Strong Model Capability} 
To overcome the dual-encoder's limitations in model capability, 
we propose a cross-encoder architecture for code search, as illustrated in Figure~\ref{intro_model}(b). 
The cross-encoder first concatenates the query and code tokens and then encodes the concatenated query and code together. 
This approach allows the self-attention mechanism in the encoder to model the token-level interactions between query and code, 
thereby improving the model's capability.

\paragraph{Balance of Effectiveness and Efficiency} 
While effective, the cross-encoder requires concatenating each query with all codes in the codebase 
and encoding all the concatenations for evaluation and online serving. 
This approach is infeasible when dealing with large codebases containing millions of code snippets. 
On the contrary, the dual-encoder architecture is efficient by encoding all codes in the codebase during data pre-processing 
and only needing to encode the query to find relevant codes based on their embeddings during online serving. 
To take advantage of both the cross-encoder's effectiveness and the dual-encoder's efficiency, 
we introduce a retriever-ranker framework for code search. 
This framework consists of a dual-encoder as the retriever module to retrieve a small set of possibly relevant codes for the query 
and a cross-encoder as the ranker module to further rank this small set of codes. 
In this way, the cross-encoder only needs to encode the query and the small amount of retrieved codes, 
making it efficient for evaluation and online serving.

\paragraph{Reasonable Training Method} 
To train the models, we sample some codes from the code corpus as negative codes (termed as negative sampling), 
and then use InfoNCE~\cite{DBLP:conf/cvpr/He0WXG20} as the optimization objective. 
Negative sampling plays an essential role in this process~\cite{DBLP:conf/iclr/RobinsonCSJ21, DBLP:conf/emnlp/LiTWFZY19}. 
The similarity scores of the well-trained dual-encoder in our retriever-ranker code search framework, 
provide reasonable signals to sample negative codes for the cross-encoder. 
The low similarity scores indicate the presence of apparent patterns that distinguish the query from these codes. 
Thus, they are less informative to train the cross-encoder in distinguishing among the codes retrieved by the dual-encoder which are hard to distinguish. 
Besides, it is impractical to consider all unlabeled codes as irrelevant (negative) since the code corpus is large and some of the unlabeled codes might actually be relevant to the query. 
We refer to these unlabeled relevant codes as false negative samples~\cite{DBLP:conf/emnlp/LiTWFZY19}, as they could be mistakenly treated as negatives without proper judgment.
We assume that unlabeled codes with extremely high similarities, as per the dual-encoder, might actually be relevant. 
If we use these codes as negative samples, it could mislead the model training. 
Therefore, to ensure a more reasonable negative sampling process, 
we rank all codes according to their similarity scores, 
remove the codes with low and extremely high rankings from the negative candidates, 
and use the left codes as negative samples to train the cross-encoder. 
This approach helps improve the training process and enhances the overall performance of the model.

The reminder of this paper is organized as below: 
Section~\ref{PRELIMINARY} introduces some preliminary knowledges about code search;
Section~\ref{R2PS} presents our method R2PS for code search; 
Section~\ref{EXP_DESIGN} exhibits the experimental design; 
Section~\ref{EVALUATION} exhibits the results to demonstrate the advantage of our method; 
Section~\ref{RELATED WORK} related work about code search and pre-trained model; 
Section~\ref{CONCLUSION} concludes this paper.

\section{PRELIMINARY}
\label{PRELIMINARY}

\subsection{Task Formulation}
Assuming we have a large code corpus containing various code snippets $c_i$, where $i=1, 2, \cdots, N$, 
each implementing a specific function. 
Given a user query $q$, described in natural language, 
the objective of code search is to quickly identify and present a small amount of relevant codes to the user based on relevance scores (also known as similarities). 
The core of code search lies in two perspectives: 
(1) Precision: precisely estimate the relevance scores of queries and codes; 
(2) Efficiency: rapidly estimate the relevance scores of the query and all codes in the code corpus. 
The definitions for used notations in this paper are shown in Table~\ref{notation}. 

\begin{table}[t!]
    \centering
    \caption{Notation and Definition.}
    \resizebox{.45\textwidth}{!}{%
    \begin{tabular}{ccccc}
    \hline
    Notation           &   Annotation     \\
    \hline
    $q_i$, $c_i$ & \begin{tabular}[c]{@{}c@{}} A query, a code snippet. In some context, they denote  \\  the token sequence of the tokenized query and code. \end{tabular} \\
    \hline
    $s^q_1,s^q_2, \cdots,s^q_l$  & \begin{tabular}[c]{@{}c@{}} The query token sequence of query $q$; \\ $l$ is the token sequence length. \end{tabular} \\
    \hline
    $s^c_1,s^c_2, \cdots,s^c_m$  & \begin{tabular}[c]{@{}c@{}} The query token sequence of query $c$; \\ $m$ is the code sequence length. \end{tabular} \\
    \hline
    $E()$ & \begin{tabular}[c]{@{}c@{}} The code pre-trained language model, \\ which is Transformer-based model.  \end{tabular} \\
    \hline
    $<, >$ & Dot product of two vectors. \\
    \hline
    $\bm Q, \bm K, \bm V$ & \begin{tabular}[c]{@{}c@{}} The query, key, value matrix in Transformer, \\ $\bm Q$ is irrelevant with user query $q$.  \end{tabular} \\
    \hline
    $\bm A$ & \begin{tabular}[c]{@{}c@{}} The attentin matrix in Transformer, \\ with $\bm A = softmax(\frac{\bm Q \bm K^T}{\sqrt d})$.  \end{tabular} \\
    \hline
    $\bm H$ & \begin{tabular}[c]{@{}c@{}} The representation matrix of token sequence.   \end{tabular} \\  
    \hline
    $[q,c]$ & \begin{tabular}[c]{@{}c@{}} The concatenation of query token sequence  \\ and code token sequence.   \end{tabular} \\
    \hline
    $s(q,c)$ & \begin{tabular}[c]{@{}c@{}} The relevance score of the query $q$ and the code $c$.  \end{tabular} \\
    \hline
    \end{tabular}%
    }
    \label{notation}
\end{table}

\subsection{Dual-encoder and Cross-encoder}
\textbf{Dual-encoder} As depicted in Figure~\ref{intro_model}(a), the dual-encoder architecture is commonly used to predict relevance scores for pre-trained model-based code search. 
Firstly, the query and code are tokenized to token sequence.  
Then, the pre-trained model encoder maps the query and code token sequence to query embedding and code embedding, respectively. 
Finally, the relevance score of the query and code is calculate by the dot product of the two embeddings. 
The dual-encoder architecture can be formulated as: 
\begin{equation}
    s_{dual}(q, c) = <E(q), E(c)>, 
    \label{dual_formular}
\end{equation}
where $E()$ denotes the pre-trained model encoder, $<, >$ denotes the dot product operation, $q$ and $c$ denote the token sequence of query and code, respectively. 
The pre-trained model is essentially a transformer-based model, with self-attention of all tokens in the sequence to learn interactions among the tokens~\cite{DBLP:conf/aaai/Hao0W021}. 

\textbf{Cross-encoder} To improve the model capability of pre-trained models for code search, 
we propose using a cross-encoder to model the token-level interactions between query and code. 
Specifically, we concatenate the query token sequence and code token sequence to a unified sequence, 
then use the pre-trained model encoder to map the unified sequence into an embedding,  
and finally use a neural network head to map the embedding to a scalar that represents the relevance score of the query and code. 
The cross-encoder architecture can be represented as: 
\begin{equation}
    s_{cross}(q, c) = NN(E([q, c])), 
\end{equation}
where $[q, c] = (s^q_1,s^q_2, \cdots,s^q_l, s^c_1,s^c_2, \cdots,s^c_m)$ denotes the concatenation of query and code, 
$E()$ denotes the pre-trained model encoder, and $NN()$ denotes the neural network head. 

In this way, the query tokens and code tokens are inputed into the Transformer model together, 
and thus all token-level interactions can be effectively modeled, including
not only interactions within query $(s^q_i, s^q_j)$ and code $(s^c_i, s^c_j)$, 
but also those between query and code $(s^q_i, s^c_j), (s^c_j, s^q_i)$. 
As a result of its ability to learn the cross-interactions between query and code, we refer to this model as the cross-encoder framework. 
Compared to the dual-encoder, the model capability of the cross-encoder is stronger, making it more precise for code search.

\section{R2PS}
\label{R2PS}

In this section, we present a retriever and ranker framework with a ranking-based hard negative sampling method for code search, 
which is a flexible and universal patch and can be applied to many code models to improve the performance of previous work for the code search task. 

\subsection{Retriever and Ranker Framework} 
During the inference stage, 
the cross-encoder must encode all possible concatenations of each query and all codes in the codebase using the encoder of pre-trained model. 
This enables the cross-encoder to retrieve the relevant code snippets based on their respective relevance scores. 
However, given that the codebase is typically vast and comprises millions of codes, 
this process demands a significant amount of computing resources. 
As a result, the cross-encoder can be slow to provide results when serving online, which can impact its efficiency.

\begin{figure*}[t]
    \centering
    \includegraphics[width=\textwidth]{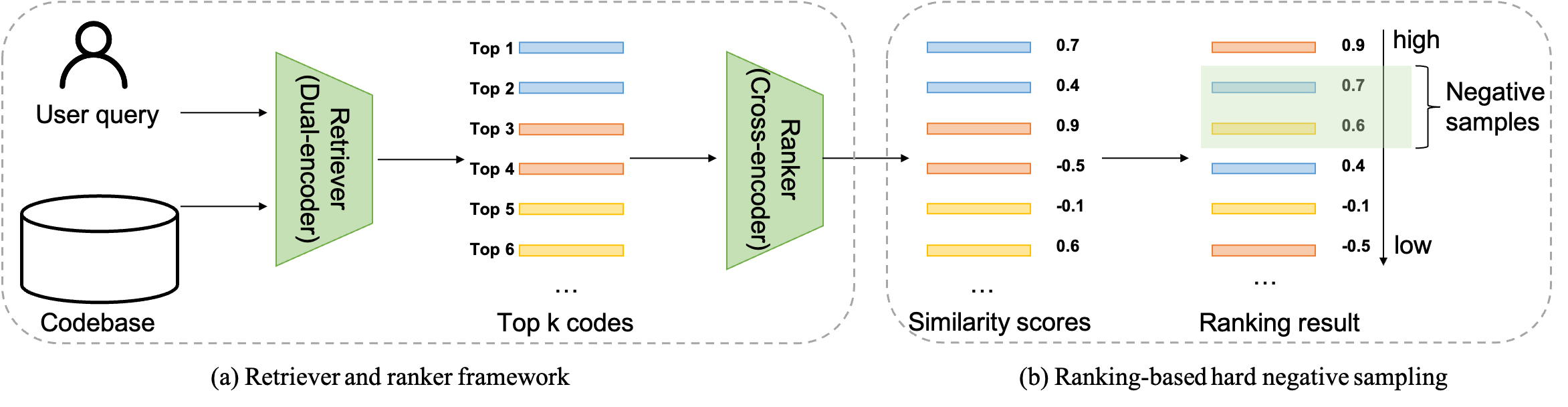}
    \caption{An overview of R2PS for code search. }
    \label{RR_frame}
\end{figure*}

To make the cross-encoder practical for code search, 
we introduce a Retriever and Ranker framework (RR) for code search, as illustrated in Figure~\ref{RR_frame} (a). 
The RR framework comprises two cascaded modules: 
(1) A dual-encoder module is employed as the retriever to identify $k$ codes with the highest relevance scores for a given user query; 
(2) A cross-encoder module is used as the ranker to rank the $k$ codes further. 

We explain the rationality behind the efficiency of the RR framework for code search as follows: 
In the dual-encoder model, we can compute the code embeddings of all the codes in the codebase during pre-processing and store them as $E(c_i)$, where $i=1, 2, \cdots, N$. 
Therefore, during evaluation and online serving, the retriever only needs to compute the embedding of the given query, $E(q)$, 
and match it with the pre-calculated code embeddings to calculate the relevance scores. 
The cross-encoder model then determines the relevance scores between the query and the retrieved $k$ codes, and ranks them accordingly. 
As a result, the online encoding time for a query only involves one forward propagation of the dual-encoder for the query 
and $k$ forward propagations of the cross-encoder for the concatenations of the query and $k$ codes. 
This computational process is independent of the number of codes in the codebase. 
By increasing $k$, the dual-encoder finds more relevant codes for the cross-encoder to rank, 
which results in better performance. 
However, with a smaller value of $k$, the cross-encoder performs fewer propagations, 
which reduces the computing cost. 
Thus, choosing the appropriate value of $k$ can balance the performance and computing cost.

\subsection{Ranking-based Hard Negative Sampling}
To train the model to enhance its ability to retrieve potentially relevant codes from the entire codebase, 
we utilize a method of negative sampling wherein we randomly select codes from the entire codebase as negative samples to train the model.
In practice, we employ the in-batch negative sampling technique, which treats the codes in the sampled batch as negative samples.
Specifically, given a batch data $\{(q_j, c_j)\}_{j=1}^{b}$, where code $c_j$ is relevant to query $q_j$, 
all other codes $c_i$, where $i \neq j$ are deemed as negative samples of query $q_j$. 
Since the batch data is randomly sampled from the training set, we can consider other codes in the batch as randomly sampled from the training set.

With a well-trained dual-encoder, we are capable of determining similarity scores between each query and all available codes. 
These similarity scores provide valuable signals for conducting negative sampling during the cross-encoder training process. 
Lower similarity scores indicate distinct patterns to differentiate the query from certain codes. 
Consequently, such codes are not informative as negative samples for training the cross-encoder.
The primary role of the cross-encoder in the retriever-ranker framework is to identify the relevant code among the potentially relevant codes retrieved by the dual-encoder. 
Therefore, to enhance the cross-encoder's ability to distinguish from the codes retrieved by the dual-encoder, 
we eliminate codes with small similarity scores from the pool of negative candidates. 
Instead, we selectively sample codes with high similarity scores as negative samples for cross-encoder training.
Besides, it is important to acknowledge that not all unlabeled codes are irrelevant to the query, given the large size of the code corpus. 
Some unlabeled codes may indeed be relevant to the query, and treating them as negative samples would be unreasonable. 
Here, we assume that codes with extremely high similarity scores are relevant to the query, 
and exclude these codes from the negative sample candidates. 
To implement the above idea, we adopt a ranking-based sampling strategy, as illustrated in Figure~\ref{RR_frame} (b). 
This involves ranking the codes based on their similarity scores of the well-trained dual-encoder 
and then selecting a small number of codes as negative samples from the relatively top-ranking positions (with top ranking but not the toppest ranking). 
This approach ensures that the cross-encoder is trained with more reasonable negative samples, 
thereby improving its ability to accurately distinguish relevant codes from the dual-encoder's retrieved codes. 
We term this method as Ranking-based Hard Negative Sampling method (PS), 
and this is actually a kind of joint training of the cascaded RR framework that trains the cross-encoder with the help of the result of the dual-encoder.

\subsection{Training Method}

The cascaded RR framework consists of two modules that perform distinct functions. 
The dual-encoder module retrieves data from the codebase, which contains all codes, 
and its ability to predict the relevance scores of all codes is crucial. 
On the other hand, the cross-encoder module ranks a small number of retrieved codes that have high relevance scores, 
as determined by the dual-encoder. 
Thus, its ability to predict the relevance scores of these potentially relevant codes is of utmost importance.

The goal of training a code search model is to learn a relevance estimation function that assigns higher scores 
to relevant codes for queries compared to those that are irrelevant. 
To achieve this, we sample some irrelevant codes as negative samples 
and aim to maximize the relevance scores of the relevant codes while minimizing the relevance scores of the sampled codes. 
Our method employs the InfoNCE loss~\cite{DBLP:journals/corr/abs-1807-03748} as loss function, 
which can be formulated as: 
\begin{equation}
    \mathcal{L}_q = - log \frac{e^{s(q,c^+)/\tau}}{e^{s(q,c^+)/\tau} + \sum_{i=1}^m e^{s(q,c^-_i)/\tau}}, 
\end{equation}
where $q$ is a query, $c^+$ is a relevant code of the query $q$, and $c_{i}^-$ is the set of sampled negative codes, 
$\tau$ denotes the temperature hyper-parameter, 
$s(q, c)$ denotes the relevance score of the query-code pair estimated by the pre-trained model encoders. 
Minimizing this loss will result in increasing the relevance score of the relevant query-code pair $s(q,c^+)$ 
and decreasing the relevance scores of the query with irrelevant codes $s(q,c^-_i)$.

\section{EXPERIMENTAL DESIGN}
\label{EXP_DESIGN}

In this section, we conduct experiments on four datasets to evaluate our proposed method. 
The experiments are designed to address the following research questions: 
\begin{itemize}[itemsep=2pt,topsep=0pt,parsep=0pt,leftmargin=*]
    \item \textbf{RQ1:} Does our proposed RR/R2PS exhibit superior performance in theory?
    \item \textbf{RQ2:} Can our proposed RR/R2PS patch boost the performance of pre-trained models for the code search task? 
    \item \textbf{RQ3:} Is the cross-encoder inefficient for online serving? If so, can our RR framework overcome this efficiency issue? 
    \item \textbf{RQ4:} How does the number of retrieved code snippets impact the performance and efficiency tradeoff?
\end{itemize}

Our proposed RR/R2PS is actually a patch during the fine-tuning stage for the code search task, 
which can be applied to various pre-trained language models designed for code. 
To fully validate the effectiveness and universality of our proposed method, we use three different code pre-trained models as backbone models of our RR/R2PS patch, 
including CodeBERT~\cite{DBLP:conf/emnlp/FengGTDFGS0LJZ20}, GraphCodeBERT~\cite{DBLP:conf/iclr/GuoRLFT0ZDSFTDC21}, and UniXcoder~\cite{DBLP:conf/acl/GuoLDW0022}.

\subsection{Datasets} 
We evaluate using four code search benchmark datasets: CodeSearchNet (CSN)~\cite{DBLP:journals/corr/abs-1909-09436, DBLP:conf/iclr/GuoRLFT0ZDSFTDC21}, 
AdvTest~\cite{DBLP:conf/nips/LuGRHSBCDJTLZSZ21}, StaQC~\cite{DBLP:conf/www/YaoWCS18}, and CoSQA~\cite{DBLP:conf/acl/HuangTSG0J0D20}. 
CSN is collected from GitHub and uses the function comments as queries and the rest as codes. 
It includes six separate datasets with various programming languages including Ruby, JavaScript, Go, Python, Java, and PHP.
AdvTest is a challenging variant of CSN-Python that anonymizes function and variable names. 
StaQC is obtained from StackOverFlow with question descriptions as queries and code snippets in answers as codes. 
CoSQA is a human-annotated dataset with queries collected from Bing search engine and candidate codes generated from a finetuend CodeBERT encoder. 
Table~\ref{Dataset_statistics} summarizes the dataset statistics, including the number of relevant query-code pairs in the training, validation, and test set, and the number of codes in the codebase used for evaluation.  
The codebase for AdvTest is comprised of all codes in the validation and test sets respectively. 

\begin{table}[t]
    \caption{Statistics of the datasets. }
    \label{Dataset_statistics}
    \resizebox{0.48\textwidth}{!}{%
    \begin{tabular}{ccccc} 
        \hline
        Dataset       & Training  & Validation & Test & Codebase \\ 
        \hline
        CSN-Ruby&24,927 & 1,261 & 1,400 & 4,360   \\
        CSN-JS & 58,025 & 3,291 & 3,885 & 13,981  \\
        CSN-Go&167,288&7,325&8,122&28,120  \\
        CSN-Python&251,820&13,914&14,918&43,827  \\
        CSN-Java&164,923&5,183&10,955&40,347    \\
        CSN-PHP&241,241&12,982&14,014&52,660    \\
        CoSQA&19,604&500&500&6,267   \\
        AdvTest&251,820&9,604&19,210&- \\
        StaQC&118,036&14,755&14,755&29,510      \\
        \hline
    \end{tabular}
    }
\end{table}

\subsection{Evaluation Metrics} 
To evaluate the performance of code search, we use Mean Reciprocal Rank (MRR), a commonly used metric for code search in previous research~\cite{DBLP:conf/acl/GuoLDW0022}. 
MRR calculates the average reciprocal rank of the relevant codes for all queries. It is defined as:
\begin{equation}
    MRR = \frac{1}{|D|} \sum_{(q,c)\, in \, D} \frac{1}{rank^{(q)}_c}, 
\end{equation}
where $rank^{(q)}_c$ is the rank of the relevant code $c$ among all code in the codebase for the query $q$, 
and $|D|$ is the total number in the dataset.

\begin{table*}[t]
    \caption{Performance comparison of different methods in terms of MRR (\%). 
    The left seven columns are the reuslts of CodeSearchNet datast. The bold is the winner in each column. 
    The numbers in the bracket are the improvement achieved by the RR/R2PS patch compared to the corresponding baseline models. }
    \label{performance_4_dataset}
    \resizebox{1\textwidth}{!}{%
    \begin{tabular}{lccccccl|lll} 
        \hline
        \multirow{2}{*}{Method} & \multicolumn{7}{c|}{CSN} & \multicolumn{1}{l}{\multirow{2}{*}{CoSQA}} & \multirow{2}{*}{StaQC} & \multirow{2}{*}{AdvTest} \\
        \cline{2-8}
        & Ruby  & JS & Go & Python & Java & PHP & Average  & \multicolumn{1}{l}{}                       &                        &    \\ 
        \hline 
        SyncoBERT     & 72.2 & 67.7 & 91.3 & 72.4 & 72.3 & 67.8    & 74.0 &  -    &   -   &  38.1    \\
        CodeRetriever &  75.3 & 69.5 & 91.6 & 73.3 & 74.0 & 68.2  & 75.3 &  69.6 &   -   &  43.0  \\
        \hline
        CodeBERT      & 68.8 & 62.7 & 89.0 & 68.5 & 68.4 & 64.2   & 70.3 &  66.9   &  21.4  &  34.9  \\
        {  }+RR  & 75.9&68.9&91.8&74.9&74.5&69.4  &  75.9\scriptsize{($\uparrow$5.6)} &  68.3\scriptsize{($\uparrow$1.4)}
           &  26.2\scriptsize{($\uparrow$4.8)}  &  42.0 \scriptsize{($\uparrow$7.1)}\\
        \hline 
        GraphCodeBERT & 69.4 & 65.4 & 90.3 & 70.8 & 70.0 & 64.1 & 71.7 &  68.7 &  20.5  &   39.1 \\
        {  }+RR     &75.8&71.8&92.5&73.6&77.1&71.0     &  77.0\scriptsize{($\uparrow$5.3)} & 69.8\scriptsize{($\uparrow$1.1)}
         & 24.3\scriptsize{($\uparrow$3.8)} & 45.4\scriptsize{($\uparrow$6.3)}  \\
        \hline
        UniXcoder     & 73.9 & 68.5 & 91.5 & 72.5 & 72.6 & 67.6   & 74.4 & 69.6  & 25.8  & 42.5  \\ 
        {  }+RR       &77.9&72.1&92.5&76.1&76.0&70.6   &  77.5\scriptsize{($\uparrow$3.1)} & 71.5\scriptsize{($\uparrow$1.9)}  
        & 29.0\scriptsize{($\uparrow$3.2)}  & 45.4\scriptsize{($\uparrow$2.9)}   \\
        {  }+R2PS         & \textbf{79.1} & \textbf{73.8}
        & \textbf{93.4} & \textbf{78.3}
        & \textbf{78.2} & \textbf{72.4}
        &  \textbf{79.2}\scriptsize{($\uparrow$4.8)} & \textbf{72.3}\scriptsize{($\uparrow$2.7)}  
        & \textbf{29.4}\scriptsize{($\uparrow$3.6)}  & \textbf{47.0}\scriptsize{($\uparrow$5.2)}  \\
        \hline
    \end{tabular}
    }
\end{table*}

\subsection{Baselines} 
The compared methods include: 
(1) CodeBERT~\cite{DBLP:conf/emnlp/FengGTDFGS0LJZ20}, which was pre-trained on the masked language modeling and replaced token detection task; 
(2) GraphCodeBERT~\cite{DBLP:conf/iclr/GuoRLFT0ZDSFTDC21}, which was pre-trained on additional graph relevant tasks such as link prediction of data flow graph; 
(3) UniXcoder~\cite{DBLP:conf/acl/GuoLDW0022}, which was pre-trained on addtional language modeling task; 
In fact, above three methods are from the same research group. 
We choose them as baselines because their work has great impact on this area, 
and is easy to follow by refering their released code and downloading their released pre-trained model. 
We modify the fine-tune setting of CodeBERT to the setting of GraphCodeBERT and UniXcoder. 
The reported results of CodeBERT, GraphCodeBERT and UniXcoder are reproduced by us by re-run their official released code 
(for CodeBERT, we modify it according to the source code of GraphCodeBERT and UniXcoder), 
and the performance is nearly the same as their reported results in their latest work of UniXcoder~\cite{DBLP:conf/acl/GuoLDW0022}. 
(4) SyncoBERT~\cite{DBLP:journals/corr/abs-2108-04556}, which was pre-trained on syntax-enhanced contrastive learning task; 
We choose this model from different research group considering the diversity. 
(5) CodeRetriever~\cite{DBLP:journals/corr/abs-2201-10866}, which was pre-trained on NL-code contrastive learning task. 
We choose this model because it is the newest pre-trained models for code search and performs well. 
The reported results of SyncoBERT and CodeRetriever are directly from their paper, 
because they have not released their pre-trained models and we cannot reproduce their results.

\subsection{Implementation Details} 
During the fine-tune stage, both the query encoder and the code encoder in the dual-encoder share parameters, 
with the same code pre-trained model used to initialize the two encoders. 
The cross-encoder is also fine-tuned based on the corresponding code pre-trained models. 
However, it does not share parameters with the previous query and code encoders in the dual-encoder framework. 
In the RR approach, we train the dual-encoder and the cross-encoder together using in-batch negative sampling strategy. 
On the other hand, the R2PS approach involves a two-step training process: 
first, we train the dual-encoder with in-batch negative sampling; 
then, we train the cross-encoder with negative sampling based on relevance scores calculated by the previous well-trained dual-encoder.

\textbf{Hyper-parameters:} We follow the same experimental settings as the series work of CodeBERT, GraphCodeBERT, and UniXcoder, 
including 10 training epochs, a learning rate of 2e-5 with linear decay, and 0 weight decay. 
We set the temperature hyper-parameter $\tau$ in the InfoNCE loss function to 0.05 for CSN and CoSQA datasets, 
and 0.025 for AdvTest and StaQC. 
The number of retrieved codes in the RR architecture is set to 10 to leverage the advantage of the cross-encoder and balance the efficiency. 
For the CoSQA dataset, we use the average of the dual-encoder and cross-encoder as the ranker module. 
In the InfoNCE loss, we set the number of negative samples to 32. 
In ranking-based hard negative sampling, 
we conduct negative sampling beginning from rank 1 by the dual-encoder for CSN, CoSQA, and AdvTest without tuning, 
and beginning from the top 0.4\% rank for StaQC. 
In fact, the default setting without tuning for most hyper-parameters can achieve satisfying performances in our method. 
However, the hyper-parameters also keep the flexibility to perform well for some rare data distributions. 
We will release our code and well-trained model when our paper is published.

\section{EVALUATION}
\label{EVALUATION}

\subsection{Theoretical Analysis (RQ1)}
We conduct some analysis to exhibit the superiority of the RR/R2PS method. 
First, we demonstrate the model capability of the cross-encoder and provide rationalization for its effectiveness in comparison to the dual-encoder. 
We then present a complexity analysis of the dual-encoder, cross-encoder, 
and RR framework to demonstrate the efficiency of the RR framework for evaluation and online serving.

\paragraph{Model Capability} 
The self-attention mechanism for the query tokens in the dual-encoder can be formualted as: 
\begin{equation}
    \bm H^{q} = \bm A^{qq} \bm V^{c}, 
\end{equation}
where $\bm A^{qq} \in \mathbb {R}^{l*l}$ denotes the attention matrix to model the interactions between all query tokens, 
$\bm V^{q}, \bm H^{q} \in \mathbb {R}^{l*d}$ is the representation matrix of all query tokens. 
Similarly, the self-attention mechanism for the code tokens in the dual-encoder can be formualted as: 
\begin{equation}
    \bm H^{c} = \bm A^{cc} \bm V^{c}, 
\end{equation}
where $\bm A^{cc} \in \mathbb {R}^{m*m}$ denotes the attention matrix to model the interactions between all token tokens, 
$\bm V^{c}, \bm H^{c} \in \mathbb {R}^{m*d}$ is the representation matrix of all code tokens. 
These two equations can be combined into one as follows: 
\begin{equation}
    \begin{pmatrix}
        \bm H^{q} \\
        \bm H^{c}
    \end{pmatrix}
    = 
    \begin{pmatrix}
        \bm A^{qq} & \bm 0 \\
        \bm 0 & \bm A^{cc} \\
    \end{pmatrix}
    \begin{pmatrix}
        \bm V^{q} \\
        \bm V^{c}
    \end{pmatrix}.
    \label{dual_matrix}
\end{equation}

The self-attention mechanism of all tokens in the cross-encoder can be formualted as: 
\begin{equation}
    \begin{pmatrix}
        \bm H^{q} \\
        \bm H^{c}
    \end{pmatrix}
    = 
    \begin{pmatrix}
        \bm A^{qq} & \bm A^{qc} \\
        \bm A^{cq} & \bm A^{cc} \\
    \end{pmatrix}
    \begin{pmatrix}
        \bm V^{q} \\
        \bm V^{c}
    \end{pmatrix}, 
    \label{cross_matrix}
\end{equation}
where $\bm A^{qc} \in \mathbb {R}^{l*m}, \bm A^{cq} \in \mathbb {R}^{m*l}$ denote the interactions between query and code tokens.

Comparing Equation (\ref{dual_matrix}) and Equation (\ref{cross_matrix}), 
we can find that the dual-encoder is actually a special case of the cross-encoder with 
$\bm A^{qc} = \bm 0$ and $\bm A^{cq} = \bm 0$. 
Hence, it is apparent that the cross-encoder possesses a stronger model capability than the dual-encoder.

\begin{figure}[t]
    \centering
    \includegraphics[width=0.43\textwidth]{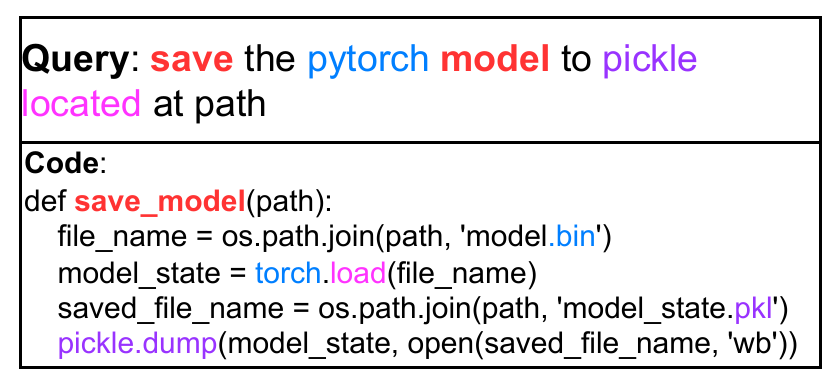}
    \caption{A case of a query and its relevant code. The same colors indicate that these tokens match between query and code. }
    \label{case_code}
\end{figure}

In Figure~\ref{case_code}, we demonstrate the importance of query-code interactions modeled by $\bm A^{qc}$ and $\bm A^{cq}$ using a case study.
To accurately predict the relevance of a query and code, it is essential to examine the various matching relationships between them in detail, 
as depicted by the same colors in the figure. 
The matching query and code tokens facilitate strong interactions in $\bm A^{qc}$ and $\bm A^{cq}$, 
which result in that their representations are enhanced by fusing information within these matching tokens each other with $\bm A^{qc}$ and $\bm A^{cq}$. 
This approach enables the model to learn the matching relationships between query and code tokens more effectively.

\paragraph{Complexity Analysis} 
Assuming that we have $M$ queries to provide them with relevant codes from a codebase containing $N$ codes, 
we analyze the complexity about the evaluation with different framework. 
The search process entails computing the relevance scores of all queries against all codes. 
These relevance scores are represented by a matrix $\bm S \in \mathbb{R}^{M*N}$, 
where each row corresponds to a query and each column represents a code.

With the dual-encoder architecture, $\bm S$ can be factorized by multiplying the embedding matrix of all queries and codes, 
as expressed in the following equation: 
\begin{equation}
    \bm S = \bm Q \bm C^T, 
\end{equation}
where $\bm Q \in \mathbb{R}^{M*d}$ and $\bm C \in \mathbb{R}^{N*d}$, $d$ denotes the embedding dimension. 
Each row in $\bm Q$ and $\bm C$ is the output of pre-trained model encoder for corresponding query and code. 
Therefore, $M$ rounds of pre-trained model propagation for queries and $N$ rounds of pre-trained model propagation for codes are sufficient. 
As a result, the dual-encoder approach offers a complexity of $\mathcal O(M+N)$. 

On the contrary, the dual-encoder cannot factorize $\bm S$ as the dual-encoder. 
Each element in $\bm S$ must be calculated by propagation of the pre-trained model encoder. 
$\bm S$ consists of $M*N$ elements, making the complexity of the cross-encoder $\mathcal O(M*N)$.

The RR framework comprises of two cascade modules. 
Firstly, a dual-encoder is used to retrieve relevant codes from the entire codebase, which has a complexity of $\mathcal O(M+N)$. 
Secondly, a cross-encoder is employed to identify relevant codes from the $k$ retrieved codes. 
For every query, the cross-encoder needs to conduct forward propagation $k$ times, 
resulting in a complexity of $\mathcal O(M*k)$ for the cross-encoder in the RR framework. 
Therefore, the total complexity of the RR framework can be expressed as $\mathcal O(M*(1+k) + N)$.

\begin{table}[t!]
    \centering
    \caption{Complexity of different code search frameworks.}
    \resizebox{.48\textwidth}{!}{%
    \begin{tabular}{ccccc}
    \hline
    Framework & Dual-encoder & Cross-encoder & RR framework \\
    \hline
    Complexity & $\mathcal O(M+N)$ &  $\mathcal O(M*N)$  & $\mathcal O(M*(1+k) + N)$ \\
    \hline
    \end{tabular}%
    }
    \label{complexity}
\end{table}

Table~\ref{complexity} summarizes the complexities of the three frameworks discussed above. 
The dual-encoder's complexity of $\mathcal O(M+N)$ makes it highly efficient to implement for evaluation. 
However, the cross-encoder's complexity of $\mathcal O(M*N)$ makes it infeasible to implement, 
as the number of queries and codes are often immense. 
In contrast, the complexity of the RR framework, given that $k$ is always a small number, 
is of the same magnitude as that of the dual-encoder but much lower than that of the cross-encoder. 
Therefore, the RR framework is also highly efficient to implement.

\subsection{Performance Comparison (RQ2)}

The performance of compared methods in terms of MRR is shown in Table~\ref{performance_4_dataset}. 
The RR patch is applied to CodeBERT, GraphCodeBERT, and UniXcoder backbone models, 
while the R2PS patch is applied only to UniXcoder, due to computing resource limitation. 
From this table, we observe the following: 
(1) Our proposed RR patch significantly improves the performance of all three backbone models, 
with an average improvement of 4.7 on CodeBERT, 4.1 on GraphCodeBERT, and 2.8 on UniXcoder across the four datasets. 
These results demonstrate the effectiveness of our proposed RR method for code search.  
(2) The performance of UniXcoder with the RR patch is better than GraphCodeBERT and CodeBERT with the RR patch, consistent with the case without the RR patch. 
This finding suggests that better-performing code pre-trained models can achieve even stronger performance when used in conjunction with the RR patch. 
(3) Overall, UniXcoder with R2PS outperforms all other methods, including the latest CodeRetriever, providing further evidence for the superiority of our proposed method. 
Besides, although we are willing to conduct experiment with the CodeRetriever as backbone model, the pre-trained model of CodeRetriever is not released. 
However, considering that CodeRetriever is better than UniXcoder and UniXcoder with R2PS outperforms the CodeRetriever significantly, 
we infer that our R2PS can improve the performance of CodeRetriever with as least the performance of UniXcoder with R2PS. 
(4) In the backbone of UniXcoder, we find that R2PS is better than the RR for about 1.3 improvement. 
Compared with RR, R2PS has additional patch of ranking-based negative sampling, 
which provides more informative data as nagative samples to train the model better. 
Such an improvement illustrates the effectiveness of our ranking-based negative sampling method for code search.

\subsection{Efficiency Evaluation (RQ3)}

\begin{table}[t]
    \caption{The response time and GPU memory comparison of the dual-encoder, the cross-encoder, and RR framework with different scales of the codebase. 
    The number on the left of ``-" is the response time (RT), and the number on the right is the GPU memory utilization (MU). }
    \label{Computing_cost}
    \resizebox{0.5\textwidth}{!}{%
    \begin{tabular}{c|cc|cc|cc}
        \hline
        \multirow{2}{*}{Method}    &  \multicolumn{2}{c|}{1000}   & \multicolumn{2}{c|}{10,000}   & \multicolumn{2}{c}{100,000}   \\
        \cline{2-7}
        & RT (ms) & MU (MB) & RT (ms) & MU (MB) & RT (ms) & MU (MB)   \\
        \hline
        Dual-encoder  &  7.3 & 1739 & 7.9 & 1739  & 17 & 1739   \\
        Cross-encoder &  4,706 & 24651 & 44,725 & OOM(55739) &  444,997 & OOM(55739)   \\
        RR         &   58 & 2471  & 61 & 2471  &   79 & 2471    \\
        \hline
    \end{tabular}
    }
\end{table}

To verify the efficiency of our proposed RR, we conducted a comprehensive comparison of the response time and GPU memory utilization of the dual-encoder, cross-encoder, and our RR in an online serving scenario. 
To simulate real-world deployment conditions, we build the deployment environment 
and measure the processing cost from the moment the server receives a user query to the completion of the relevant code inference. 
Each user query is fed individually to the server, without batching operations, 
whereas the codes associated with a query were fed using batching operations. 
Notably, the code embeddings in both the dual-encoder and RR are pre-calculated during the pre-processing stage, 
ensuring that the response time analysis excludes the time required for code embedding in the dual-encoder. 
Our experiments are conducted on an A100 GPU, 
employing codebase scales of 1,000, 10,000, and 100,000 codes, 
and we measure the average response time for 100 query requests.

The result is shown in Table~\ref{Computing_cost}. 
The full batch processing in the cross-encoder can handle the codebase of 10,000 codes, 
but attempting to process 10,000 or 100,000 codes within a batch will lead to out-of-memory (OOM) errors on an 80G A100 GPU. 
To address this issue, we opted for a batch size of 2,500 for 10,000 codes and 100,000 for the cross-encoder to solve the problem. 
The observations from Table V are as follows:
(1) For the three scales of the codebase, 
both the dual-encoder and the RR exhibit efficient performance for online serving, 
with a cost time of no more than 100ms and GPU memory usage not exceeding 3G. 
This demonstrates the effectiveness of both methods in terms of response time and GPU memory utilization.
(2) In contrast, the cross-encoder exhibits significantly longer cost times and higher GPU memory consumption, 
especially for the codebase with 100,000 codes, which exceeds 7 minutes and utilizes 55G of GPU memory. 
Consequently, the cross-encoder is impractical for real-world applications due to its resource demands.
(3) As the codebase increases by 10 times, the time taken by the cross-encoder increases by a factor of 10, 
and the GPU memory requirements for the cross-encoder would be the same of 10 times increasing without batching operations. 
But the computational time and GPU memory requirements for the dual-encoder and RR change little. 
This is due to the fact that the cross-encoder necessitates encoding the concatenation of all queries and codes in the codebase during online serving, 
while the dual-encoder only requires encoding queries, and the RR needs to encode queries and the concatenated queries with a small number of retrieved codes during online serving. 
The additional time for the dual-encoder and RR for a larger codebase originates from ranking more scores of codes within the dual-encoder.

\subsection{Hyper-parameter Analysis (RQ4)}

\begin{figure*}[t]
    \centering
    \includegraphics[width=0.95\textwidth]{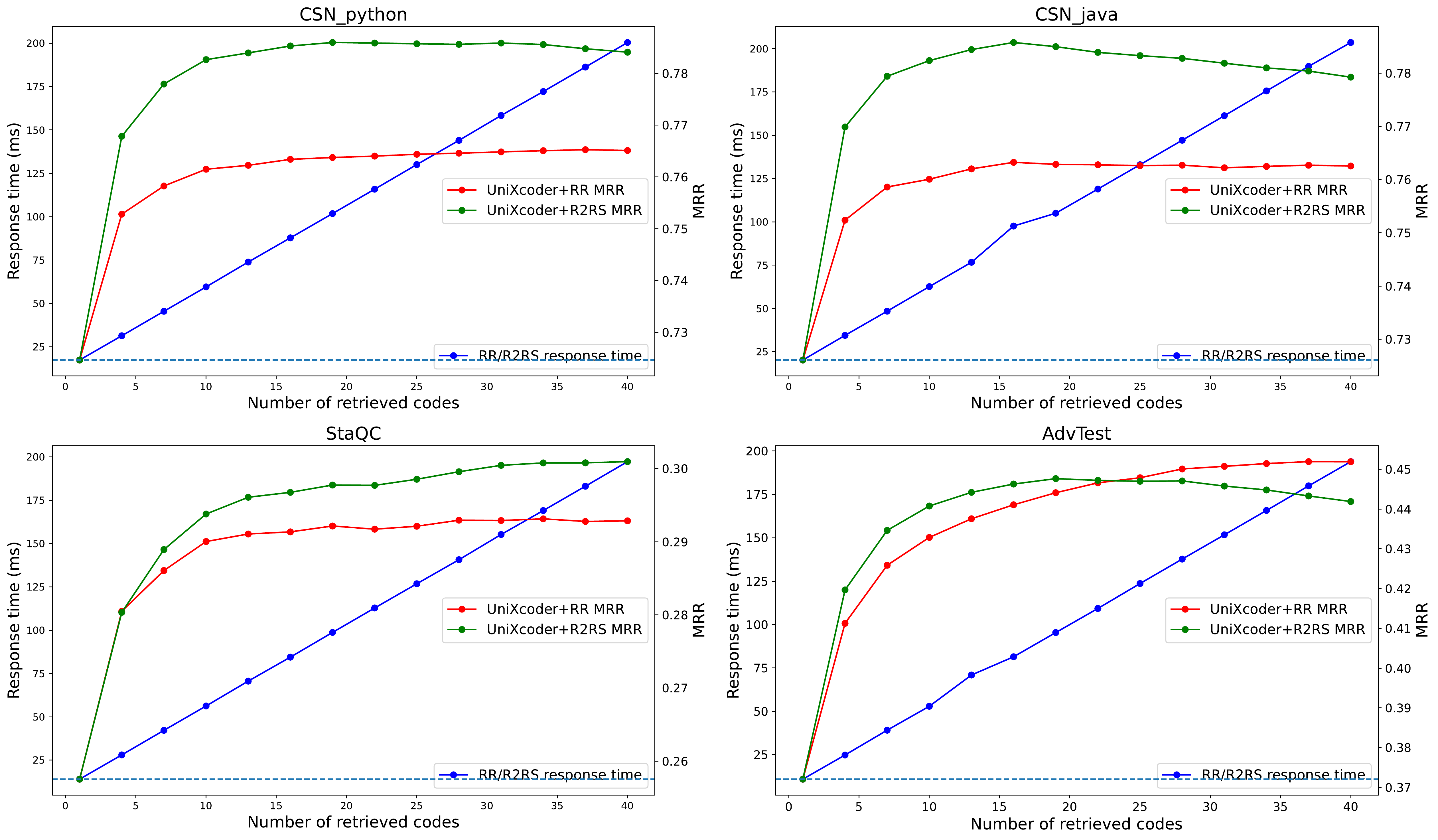}
    \caption{Performance and response time of the RR and R2PS patched UniXcoder with the different number of retrieved code $k$. 
    The below dashed line is the performance of UniXcoder without RR or R2PS patch. }
    \label{balance}
\end{figure*}

We conducted experiments to investigate the impact of varying the number of retrieved codes, $k$, on both the performance and response time during online serving of our framework. 
Prolonged response times can lead users to experience impatience while waiting.
From the result shown in Figure~\ref{balance}. we can observe that: 
(1) In general, the MRR of RR/R2PS increases with the larger number of retrieved codes $k$, 
but the rate of increase slows down with higher $k$ values. 
This suggests that retrieving more codes can enhance performance, but the improvement becomes less significant as more codes are retrieved. 
(2) The performance of R2PS is better than RR in most cases with different $k$ and different datasets, 
demonstrating the effectiveness of our proposed ranking-based hard negative sampling strategy for code search. 
This strategy helps in selecting better codes for training the model, resulting in improved performance. 
(3) Response time increases linearly as $k$ increases due to more forward propagation required by the cross-encoder. 
To reduce the response time, it is better to use a smaller value of $k$. 
(4) The performance of R2PS declines slightly in the CSN-python, CSN-Java, and AdvTest datasets with large $k$ values. 
With fewer retrieved codes, the candidate codes for the cross-encoder are more difficult, 
whereas with more retrieved codes, the candidate codes are easier. 
While hard negative sampling in R2PS enhances the model's ability to handle hard negative samples, 
its ability to distinguish easier codes may decrease. This may explain the decrease in R2PS performance with large $k$ values.
(5) Setting $k$ less than 10 causes a significant decline in performance, while $k$ greater than 20 causes a relatively prolonged response time of over 100ms. 
Therefore, we recommend setting $k$ between 10 and 20 in our RR/R2PS method.

\section{RELATED WORK}
\label{RELATED WORK}

\subsection{Code Search}

Code search aims to find relevant codes for given queries~\cite{DBLP:journals/tacl/LuanETC21, DBLP:conf/pldi/SachdevLLKS018}. 
The core of code search is to predict whether queries are relevant to codes~\cite{DBLP:conf/icse/GuZ018, DBLP:conf/sigir/BuiYJ21}. 
To determine the relevance score, traditional information retrieval (IR) methods mainly rely on rule-based code search, 
such as term matching (e.g., term frequency-inverse document frequency, TF-IDF), 
and manually-designed features based on the analysis of the abstract syntax tree of codes~\cite{DBLP:journals/pacmpl/LuanYBS019, DBLP:conf/msr/SismanK13, DBLP:conf/icse/DiamantopoulosK18}. 
However, these methods lack the ability to understand semantic information and have some limitations: 
(1) The term mismatch problem limits the ability to identify potentially useful codes; 
(2) Manually-designed features are inadequate for capturing complex and implicit features.

To overcome the limitations of IR-based methods, deep learning has been applied to the code search task to learn semantic representations of queries and codes~\cite{DBLP:conf/pldi/SachdevLLKS018}. 
The relevance score is typically calculated by the query embedding and the code embedding by some operations such as dot-product~\cite{DBLP:conf/sigsoft/CambroneroLKS019}. 
The query embedding is generated using natural language processing (NLP) models such as fastText~\cite{DBLP:conf/pldi/SachdevLLKS018}, 
LSTM~\cite{DBLP:conf/icse/GuZ018}, or GRU~\cite{DBLP:conf/sigsoft/CambroneroLKS019}. 
Code can be represented by various data structures, such as token sequence, 
Abstract Syntax Tree (AST)~\cite{DBLP:journals/tkdd/LingWWPMXLWJ21} through code static analysis~\cite{DBLP:journals/pacmpl/LuanYBS019}, variable and function name~\cite{DBLP:conf/icse/GuZ018}. 
To encode these different structures, a variety of models have been proposed~\cite{DBLP:conf/kbse/WanSSXZ0Y19}, 
including LSTM~\cite{DBLP:conf/kbse/WanSSXZ0Y19}, 
MLP~\cite{DBLP:conf/icse/GuZ018}, and Graph Neural Network~\cite{DBLP:journals/tkdd/LingWWPMXLWJ21}.  
While these techniques reduce the need for human-designed rules, 
they still require human effort to create effective data structures and suitable models to represent to code.

\subsection{Code Intelligence Pre-trained Models} 
Motivated by the huge succees of pre-trained models in the NLP field, 
many researchers introduce pre-trained models to code intelligence field~\cite{DBLP:journals/corr/abs-2201-10866, DBLP:conf/icse/NiuL0GH022, DBLP:conf/kbse/ZhangP0LG22, DBLP:conf/naacl/WangWWWZLWL22, DBLP:conf/nips/LachauxRSL21}. 
Various code pre-trained models have been trained from different perspectives, 
with some based on language modeling~\cite{DBLP:conf/emnlp/FengGTDFGS0LJZ20, DBLP:conf/naacl/AhmadCRC21}, 
and others based on code-relevant tasks such as link prediction on abstract syntax tree (AST) and data flow graph (DFG)~\cite{DBLP:conf/iclr/GuoRLFT0ZDSFTDC21, DBLP:conf/icse/NiuL0GH022}, 
or contrastive learning with regard to multi-modals about code~\cite{DBLP:conf/acl/GuoLDW0022}. 
Following pre-training, code pre-trained models can be fine-tuned on code downstream tasks, such as code search and code generation, 
which significantly outperform previous models.

While much of the research on code pre-trained models focuses on designing different pre-training tasks to improve their general ability, 
little attention has been given to improving the fine-tuning stage for specific downstream tasks. 
This suggests that the potential of pre-trained models for downstream tasks has not been fully exploited. 
In this paper, we focus on improving the fine-tuning stage of pre-trained models for the code search task, 
exploring ways to make the models more suitable for this task.

\subsection{Fine-tuning for Code Search} 
When using pre-trained models for code search, 
the most commonly employed approach is the dual-encoder architecture~\cite{DBLP:conf/iclr/GuoRLFT0ZDSFTDC21, DBLP:conf/acl/GuoLDW0022, DBLP:conf/naacl/WangWWWZLWL22}. 
In this method, the pre-trained model serves as an encoder to extract query and code embeddings, which are then used to compute the relevance score. 
Notably, the dual-encoder architecture for fine-tuning code pre-trained models is similar to prior deep learning approaches for training models on code search tasks~\cite{DBLP:conf/icse/GuZ018}. 
The key difference lies in the choice of encoder. 
While previous work utilized encoders such as LSTM and GNN, the pre-trained model based models relies on the pre-trained Transformer encoder. 
Recently, Gotmare et al.~\cite{DBLP:journals/corr/abs-2110-07811} and Hu et al.~\cite{DBLP:conf/wsdm/HuWD00H023} proposed to cascade the dual-encoder and cross-encoder for code search. 
However, the two work trains the cross-encoder by the binary classification objective without utilization of the well-trained dual-encoder. 
In contrast, we train the cross-encoder with InfoNCE objective, the loss which is more suitable for the search task, 
by the utilization of the well-trained dual-encoder, 
which can be regarded as a kind of joint training of the dual-encoder and cross-encoder for code search.

To optimize the model for the code search task, various loss functions have been used. 
One method treats the task as binary classification, determining if a query is relevant to a code, 
using a binary classification loss as proposed in~\cite{DBLP:conf/acl/HuangTSG0J0D20}. 
However, code search is essentially a ranking task, requiring all codes to be ranked by relevance to a query. 
Therefore, binary classification loss is not fully aligned with this goal. 
To address this, 
some previous works have proposed using pairwise loss 
to optimize the model by maximizing the relevance score margin between relevant and irrelevant codes for a query~\cite{DBLP:conf/icse/GuZ018, DBLP:journals/tkdd/LingWWPMXLWJ21, DBLP:journals/tosem/ZengYLXWGBDL23, DBLP:conf/kbse/WanSSXZ0Y19}. 
Furthermore, other studies suggest optimizing the model by maximizing relevance scores for relevant codes while minimizing them for irrelevant codes~\cite{DBLP:journals/corr/abs-2201-10866, DBLP:conf/iclr/GuoRLFT0ZDSFTDC21, DBLP:conf/emnlp/LiMLHHZW22}.

\section{CONCLUSION}
\label{CONCLUSION}

In this paper, our aim is to enhance the performance of pre-trained models for the code search task during the fine-tune stage. 
To achieve this, we introduce several novel approaches. Firstly, we propose a cross-encoder for code search, 
which enhances the model's capabilities compared to the typical dual-encoder. 
Next, we propose a retriever and ranker framework for code search that balances both effectiveness and efficiency. 
Finally, we propose a ranking-based negative sampling method to train the retriever and ranker framework better. 
We conducted thorough experiments and found that our R2PS system significantly improves performance while incurring an acceptable extra computing cost.

In light of our findings, we propose two potential future directions for research. 
Firstly, while the ranking-based negative sampling (PS) method we used for negative sampling proved to be effective in our experiments, 
it was relatively simple as it is only based the relevance scores of the dual-encoder. 
We suggest exploring more sophisticated and reasonable sampling methods to further enhance the model's performance. 
Secondly, while our paper focused solely on the code search task for code pre-trained models, 
these code pre-trained models can also be used for other downstream tasks such as code generation and bug fixing. 
We believe that exploring fine-tune methods for code pre-trained models in other downstream tasks would be a meaningful area of research.

\bibliographystyle{IEEEtran}
\bibliography{references}{}
\vspace{12pt}
\color{red}

\end{document}